\documentclass[11pt]{article}

\usepackage{graphicx}

\begin{document}

\title{Relationship between sunspot number and total annual precipitation 
at Iza\~na (Tenerife): \\
Maximum precipitation prediction with three year 
lagged sunspots?}

\author{Xavier Calbet, Mar{\'\i}a Carmen Romero,\\
 Juan Manuel Sancho, Pilar R{\'\i}podas \\
and
V{\'\i}ctor Jes\'us Quintero \\
C/Alfonso el Batallador, N. 1, 4--C,\\
E--31007 Pamplona,\\
Navarra,\\
Spain.\\
e-mail: xca@ll.iac.es}

\maketitle

\begin{abstract}
A possible relationship between sunspot number and total annual
precipitation from the Iza\~na Observatory has been found.
The annual precipitation period ranges from 1916 to 1998, thus
including nearly eight 11--year solar cycles.

When points of 
total precipitation for a given year at Iza\~na
are plotted on the ordinate axis versus the yearly sunspot number
on the abcisa axis three years back from the precipitation one,
nearly all of them lie in the lower left hand corner of the diagram.
This seems to indicate
a relationship between the above mentioned variables.

If this relationship is confirmed it would permit the prediction
of a maximum annual precipitation at Iza\~na three years in
advance.

\end{abstract}

\section{Introduction}

Many relationships between sunspot number and precipitation
have been found. The most simple one plotting sunspot number,
or a similar related quantity,
and some direct or indirect measurement of precipitation
versus the year when the events took place.
Sometimes one of the plotted variables is lagged several years. 
A visual correlation
or anti--correlation is normally found. Some examples
are the controversial water levels of lake Victoria \cite{brooks1923},
the lake Michigan water levels \cite{wilson1946},
the Mississippi river flow \cite{perry1996},
the drought cycles in the western United States
\cite{mitchell1979} or the also controversial Indian cyclones
\cite{meldrum1872}.
Another technique is to calculate the power
spectrum of the time series with sensitive methods
like maximum entropy methods. Some examples
are the summer Nile floods \cite{currie1987}
and the Atlantic tropical cyclones \cite{cohen1975}.
For more sun--climate relationships see the excellent reviews
from V\'azquez \cite{vazquez1998}, Hoyt and Schatten \cite{hoyt1997} 
and Pittock \cite{pittock1983}.

More recent and closer examples can be found in 
the work of Jonsson et al. \cite{jonsson2000},
where they analyze the Iza\~na precipitation and the tree rings
of the Canarian pines and find that during long drought periods
there is a strong minima in tree ring growth, but more conspicuous
minima happen during a phase of maximum solar activity.
Alves \cite{alves2000} finds a correlation between yearly sunspot number
and precipitation in Portugal,
which is very strongly influenced by Atlantic meteorological perturbations
like is also frequently the case in the Canary islands.

In this paper we will show a different type of relationship
between precipitation and sunspots. 
Its main feature is that if it is confirmed, it will permit
some type of long term precipitation prediction.
We will first present the
different datasets used in this paper and show their relationship.
A statistical test will be made on the sample to see how well
the relationship holds. Finally, if the relationship is really
produced by cause and effect 
we will show how precipitation predictions can be made and
will point some possible
physical factors behind it.

\section{The precipitation dataset}

\begin{figure}[t]
\vspace{0.5cm}
\begin{center}
\includegraphics[angle=270,width=0.9\textwidth]{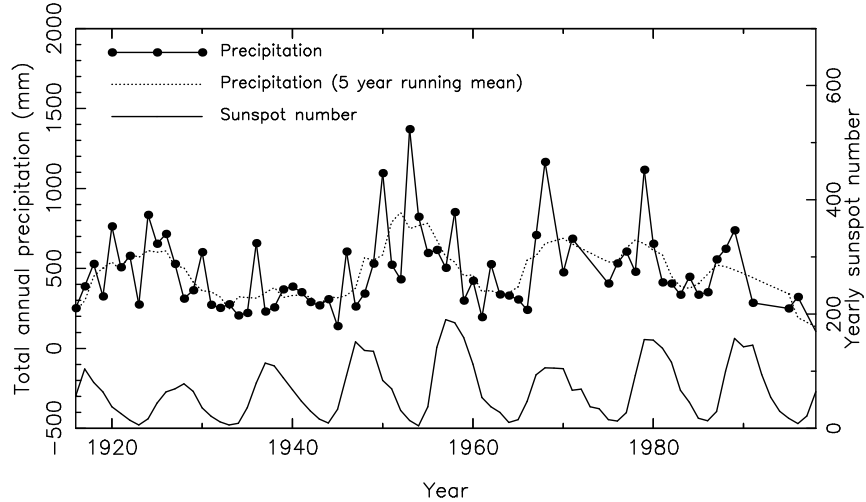}
 \caption{
Total annual precipitation series at Iza\~na,
shown with solid lines and big points. Over it
is drawn the 5 year running mean with a dotted line
for illustrative purposes.
At the bottom is shown the yearly sunspot number with a solid
line.}
\label{fig:series}
\end{center}
\end{figure}

The data was obtained from 
the local
West Canaries Meteorological Center
(``Centro Meteorol\'ogico Territorial de Canarias
Occidental'')
which belongs to the Spanish Meteorological Institute
(``Instituto Nacional de Meteorolog{\'\i}a'').
The time series consists
of the total annual precipitation registered from
1916 to 1998 inclusive at the Meteorological Iza\~na Observatory.

The Meteorological Iza\~na Observatory is located
at $28^\circ 18'{\rm N}$, $16^\circ 29'{\rm W}$ on the island
of Tenerife (Canary Islands) at 2367 m above sea level.
There is generally a strong and persistent subsidence
inversion over this region with the top located at about 1200 m
in summer and about 1800 m in winter. The Observatory
normally lies above the top of the inversion layer
although it sometimes reaches the station under certain
synoptic conditions. This is why Iza\~na is considered
a very good station for representative atmospheric
measurements of the free troposphere.

The time series of basic meteorological parameters
from this Observatory began in the 1916 when
the Iza\~na Observatory was founded. Since then the
keen work of the Observers has provided us with such a
long record.
Although the precipitation is a difficult parameter to measure
because it is not easy to devise an instrument which will
collect all the rainfall with few errors, it is
not severely affected by the time when the measurement
of the total collected precipitation is done, neither is it
very sensitive to other local conditions like temperature,
humidity, etc.
Note that in the precipitation series not all years from
1916 to the present are available. The reason for this
is that some years are not yet available in digital format.

\section{The sunspot dataset}

The yearly mean sunspot number, $R_i$, is obtained from
the United States National Geophysics Data Center
(NGDC)
at http://web.ngdc.noaa.gov
which consists of the yearly mean of the well known
International Sunspot Number.

Briefly, the International Sunspot Number, $R_i$,
results from a statistical treatment of the data 
originating from more than twenty--five observing stations.
Normally sunspots appear isolated or forming groups.
Each isolated cluster of sunspots is termed a sunspot group,
and it may consist of one or a large number
of distinct spots.
The relative sunspot number is defined as 
$R = K (10g + s)$, where $g$ 
is the number 
of sunspot groups and $s$ is the total number of distinct spots.  The scale 
factor $K$ (usually less than unity) 
depends on the observer and is intended to bring all measurements
to the same scale.

\section{Relationship between total annual precipitation
and yearly sunspot number}

\begin{figure}[t]
\begin{center}
\includegraphics[angle=270,width=0.9\textwidth]{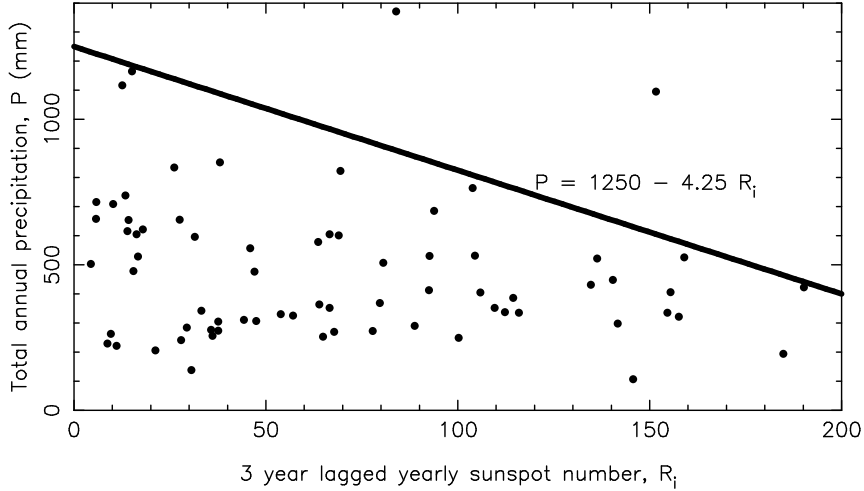}
\caption{
Three year lagged yearly sunspot number, $R_i$, versus total annual
precipitation, $P$, at Iza\~na. Also shown, as a line, is the place
where a clear cut in the points density is observed. Above
that line, $P = 1250 - 4.25 R_i$ there are very few points.}
\label{fig:triangle}
\end{center}
\end{figure}

The total annual precipitation and the yearly sunspot number 
time series is shown
in Figure \ref{fig:series} along with the 5 year
running mean for the precipitation. There does not seem to exist
any clear relationship between the two variables.
But if 
the three year lagged yearly sunspot number
versus the total annal precipitation at Iza\~na is plotted, we obtain
Figure \ref{fig:triangle}. It can be seen that the points
concentrate on the lower left hand corner of the graph.
Even more, we can plot the line $P = 1250 - 4.25 R_i$, 
which is also
shown in the figure, below which nearly all
points lie.

\begin{figure}[t]

\begin{center}
\includegraphics[angle=270,width=0.9\textwidth]{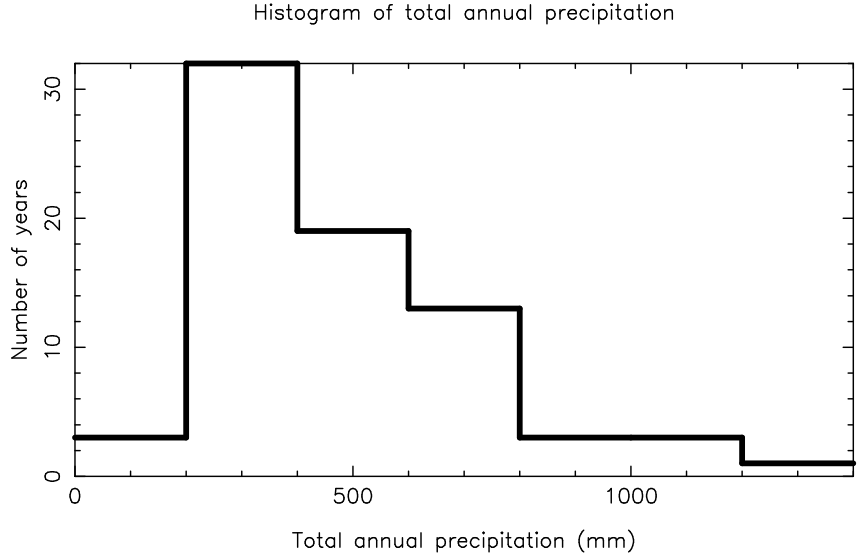}
\caption{
Total annual precipitation histogram.}
\label{fig:hpre}
\end{center}
\end{figure}

\begin{figure}[t]
\begin{center}

\includegraphics[angle=270,width=0.9\textwidth]{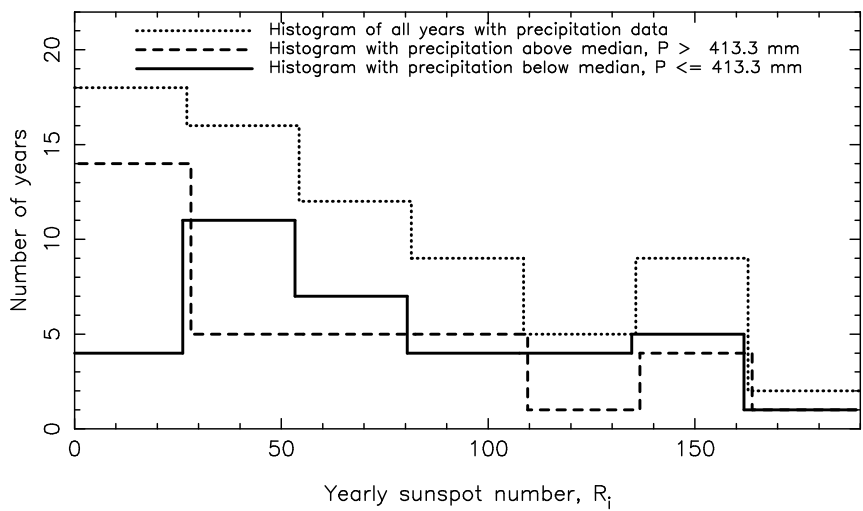}
\caption{
Histograms of yearly sunspot number, lagged three years with
respect to precipitation.
With a dotted line is shown the histogram with all years
with available precipitation data.
The dashed line shows the histogram of data points above
the median total annual precipitation, $413.3 \ {\rm mm}$.
The solid line shows the histogram of data points below
the median total annual precipitation, $413.3 \ {\rm mm}$.}
\label{fig:hspo}
\end{center}
\end{figure}

In a sense it is normal to expect this distribution
of points if we assume that both
variables are totally independent 
since most values of precipitation
and sunspot number are in the lower range of all
possible values. The histogram of
total annual precipitation is shown in 
Figure \ref{fig:hpre},
its range is from $0$ to $1400 \ {\rm mm}$
 and peaks
at around $300 \ {\rm mm}$. The histogram of yearly sunspot number
is shown in Figure \ref{fig:hspo}, its range is
from $0$ to $190$ and peaks at around 
$R_i = 20$.

A statistical test can be performed on the data to check whether
this distribution is due to chance or not.
The dataset is divided in two samples by selecting
points with a total annual precipitation higher 
and lower than its median precipitation.
If precipitation and three year lagged sunspot number are totally independent
variables then these two samples should be homogeneous,
that is, the sunspot number histogram from each of the two samples
should be the same
in a statistical sense.
In Figure~\ref{fig:hspo} the three year lagged 
sunspot number histograms are shown,
both for points with precipitation above the median, $413.2 \ {\rm mm}$,
and below it.
The actual numbers from these histograms are shown in
Table \ref{tab:years}.
Note that
the last four histogram intervals of Figure \ref{fig:hspo}
have been grouped into one 
in Table 1
so that it is statistically significant.

\begin{table}
\begin{center}
\caption{Number of years with total annual precipitation,
$P$, above and below
median, $413.3 \ {\rm mm}$, in each yearly sunspot number,
$R_i$, interval. Note that
the last four histogram intervals of Figure~\ref{fig:hspo}
have been grouped into one 
so that it is statistically significant.} 
\vspace{5pt}
\begin{tabular}{crrrrr} 
\hline
$R_i$ & & & & & Total \\
interval 
&  0--27 & 27--54 & 54--81 & 81--190 & $n_i$ \\
\hline
Years with \\
$P > 413.3 \ {\rm mm}$ &
14  &  5  &  5  &  11  & 35 \\
Years with \\
$P \le 413.3 \  {\rm mm}$ &
4  &  11  &  7  &  14  & 36\\
\hline
Total  number \\
of years \\
$m_j$ &  
18  &  16  &  12  &  25 & 71\\
\hline

\end{tabular}
\label{tab:years}
\end{center}
\end{table}

Let us denote the sample number with $i$, differentiating
data points with precipitation above and below the median; the class
number with $j$, indicating the histogram interval
of lagged sunspot number;  the number of years
in a specified interval and sample with $n_{ij}$;  the total number
of years under each sample with $n_i$;  the total number of years
under each class with $m_j$; and the total number of years
with $n$. We can make a chi--square test to test the null
hypothesis, $H_0$, that both samples are homogeneous by calculating
the parameter

\begin{equation}
\lambda = \sum_{i=1}^{2} \sum_{j=1}^{4} 
\frac{( n_{ij} - n_{i} m_{j} / n )^{2}}{n_i m_j / n}
\end{equation}

which should follow a chi--square distribution with
three degrees of freedom.
Calculating $\lambda$ with the values from Table 1
we get

\begin{equation}
\lambda = 8.49 .
\end{equation}

If we compare this values with the chi--square value
for a $5\%$ significance level,

\begin{equation}
\chi^2_{3;0.05} = 7.815
\end{equation}

we find that $\lambda > \chi^2_{3;0.05}$ and we can 
conclude that the null hypothesis
is false. In other words,
if we assume both samples are really homogeneous there
is less than $5\%$ probability that the observed
samples are produced by chance.
We thus conclude, with a high degree of confidence,
that both samples are not independent and there is
effectively a relationship between both variables.

\section{Maximum precipitation prediction with sunspot number}

If we believe that there is really a cause and effect relationship
between both variables, predicting the maximum precipitation for
Iza\~na is straightforward. We first get the value of yearly
sunspot number for a given year, $R_i$. Three years later, the maximum
precipitation at Iza\~na most probably will not exceed the value
given by the line shown on Figure~\ref{fig:triangle},
$P = 1250 - 4.25 R_i$.

\section{Discussion and possible physical causes}

As in any statistical relationship there can be many
errors along the way, so this result should be taken with
care. The measurement of precipitation is not an easy task
and is prone to errors. Depending on the wind and the position
of the pluviometer the collected water can be very different
from the actual rainfall. This is even more complicated when
the precipitation consists of snow.
Even though a relationship seems to exist, this does not
necessarily mean they form a true cause and effect relationship.
The fact that the relationship is found at Iza\~na
might be due to the fact that it is an
atmospheric background site due to its position
and altitude.
In any case, study of other meteorological data should be done to
strengthen or diminish such relationship.

Currently, in the literature, there are two possible physical
links between sunspot number and some terrestrial variable
which lags the sunspot number in about three years.

One of such physical links is the 
influence on the terrestrial magnetic field
by solar activity. It is well known that the maximum value
for the aa geomagnetic index usually follows sunspot maximum
by two or more years. In the same way, the aurorae follow
by about three years the sunspot maximum.
It has been proposed that the solar activity modulates the
cosmic rays, which in turn affect terrestrial climate
\cite{tinsley1989, svensmark1997}.
Recently \cite{svensmark2000} it has been noted that there
is a significant correlation
of magnetic activity with low level clouds ($< 3 {\rm Km}$).

Another physical link, which is most often cited in the
literature, is the different heating effect due to changes in solar
irradiance with Sun activity.
An example is the correlation of sea surface temperature
with solar irradiance \cite{reid1991}, which also has
an approximate lag of three years.
Recently \cite{white2000}
proposed a form of solar irradiance excitation of Earth's climate.

\section{Acknowledgments}
We are grateful to the Observers of the
Iza\~na Observatory for the dataset they
have provided through the years.

\end{document}